\documentclass[preprint,12pt]{elsarticle}

\usepackage{amssymb}
\usepackage{attachfile}
\usepackage{pdfpages}

\usepackage[utf8]{inputenc}     
\usepackage{amsmath,amsfonts,amsthm,amssymb,isomath}
\usepackage{graphicx}
\usepackage{indentfirst}
\usepackage{natbib}
\usepackage{caption}
\usepackage{subcaption}
\usepackage{tikz}
\usetikzlibrary{spy,calc}
\usepackage{hyperref}
\usepackage{lipsum}
\usepackage[htt]{hyphenat}
\usepackage[indent]{parskip}
 \usepackage{setspace}
\usepackage{ulem} 
\usepackage{tabularx}
  \newcolumntype{L}{>{\raggedright\arraybackslash}X}

\hypersetup{
	colorlinks=true,
	linkcolor=black,
	filecolor=blue,
	citecolor = black,      
	urlcolor=blue
}
\usepackage[makeroom]{cancel}

\usepackage{eucal}
\usepackage{upgreek}
\usepackage{xurl}

\usepackage[T1]{fontenc}

\makeatletter
\def\ps@pprintTitle{%
  \let\@oddhead\@empty
  \let\@evenhead\@empty
  \let\@oddfoot\@empty
  \let\@evenfoot\@oddfoot
}
\makeatother

\begin{document}

\begin{frontmatter}



\title{Radiation conduction in polaritonic nanowires}


\author[label1]{L\'{i}via C. McCormack}
\author[label1]{Mathieu Francoeur}
\ead{mfrancoeur@mech.utah.edu}
\affiliation[label1]{organization={Department of Mechanical Engineering},
            addressline={The University of Utah}, 
            city={Salt Lake City},
            postcode={84112}, 
            state={UT},
            country={USA}}

\begin{abstract}
Phonon polaritons have attracted increasing interest as a means to offset the reduction in phonon thermal conductivity in nanostructures caused by enhanced boundary scattering. However, the interpretation of the limited experimental data on phonon polariton-mediated conduction is hindered by the lack of comprehensive full-wave models for predicting thermal electromagnetic transport in solid-state systems. Here, the radiative thermal conductivity in the diffusive regime along SiO$_2$ nanowires near room temperature is predicted using the fluctuational electrodynamics-based discrete system Green’s function method. At 400 K, for nanowire diameters of 66 nm and 132 nm, modest radiative conductivities of 0.0264 W m$^{-1}$ K$^{-1}$ and 0.0165 W m$^{-1}$ K$^{-1}$, respectively, are obtained, with contributions dominated by surface phonon polaritons. In contrast, the kinetic theory combined with nanowire dispersion relations derived from Maxwell's equations predicts conductivities that can be nearly two orders of magnitude larger than those obtained from fluctuational electrodynamics, and are largely dominated by bulk phonon polaritons propagating within the nanowire volume. Improved agreement between the kinetic theory and fluctuational electrodynamics for the total radiative conductivity can be achieved by introducing an effective mean free path that accounts for the reduced mean free path of bulk phonon polaritons outside the Reststrahlen spectral bands of SiO$_2$. However, even with this correction, the kinetic theory fails to accurately capture the spectral distribution of radiative conductivity. This work establishes a solid foundation for the development of phonon polariton-based systems for thermal management in micro/nanoelectronic devices.
\end{abstract}




\end{frontmatter}



Phonon polaritons arise from the coupling between electromagnetic waves and transverse optical phonons in polar dielectric materials such as SiO$_2$, SiC, and SiN. In particular, surface phonon polaritons (SPhPs), which propagate along the interfaces of polaritonic and dielectric materials and are characterized by evanescent electromagnetic fields decaying normal to those interfaces \cite{joulain2005surface}, have emerged as a powerful mechanism for engineering nanoscale light-matter interactions. For example, SPhPs have been extensively exploited to enhance near-field radiative heat transfer beyond the blackbody limit \cite{rousseau2009radiative,shen2009surface,st2016near,kim2015radiative,ghashami2018precision,tang2020near,tang2024corner}, as well as to control spectral and directional thermal emission \cite{greffet2002coherent}. 

Recently, SPhPs have been investigated in the context of thermal transport in solid-state systems \cite{tranchant2019two,wu2020enhanced,wu2022observation,pei2023low,pan2023remarkable,li2024long,li2025quasi,sun2025abnormal}, primarily as a means to compensate for the reduction in phonon thermal conductivity in dielectric nanostructures \cite{chen2005surface,li2024phonon,volz2024heat}.  For example, a radiative thermal conductivity as high as 5.8 W m$^{-1}$ K$^{-1}$  at room temperature has been reported in experiments on 3C-SiC nanowires with Au-coated ends \cite{pan2023remarkable}. In addition, experiments have shown that hyperbolic phonon polaritons can contribute to the thermal conductivity of bulk anisotropic materials at levels comparable to, or even exceeding, those of phonons \cite{salihoglu2020energy,chen2024greatly}.

Thermal electromagnetic transport can be modeled using fluctuational electrodynamics (FE), a framework in which fluctuating currents that generate thermal electromagnetic fields are incorporated into Maxwell's equations \cite{rytov1989principles,greffet2026unified}. Although FE has been applied for decades to simulate  near-field radiative heat transfer \cite{polder1971theory,loomis1994theory}, radiation conduction is typically predicted using the kinetic theory (i.e., the Boltzmann transport equation) combined with interface dispersion relations derived from Maxwell's equations \cite{chen2005surface,ordonez2013anomalous,ordonez2014quantized,ordonez2014effects,tachikawa2020high,guo2021heat,tachikawa2022plane,yun2022modeling,minyard2023material,kim2023boosting,kim2023plasmon,li2023long,cheng2024temperature,yun2024maximum,jafari2026asymmetric,lv2026quantized,yun2026impact}. However, the kinetic theory does not account for coherent transport and often lacks a rigorous treatment of the system boundaries. 

To date, there is no consensus on the validity of the kinetic theory for modeling radiation conduction, and only a limited number of studies have evaluated its accuracy through comparisons with FE results in nanofilms \cite{chen2010heat,kruger2024scale} and arrays of spaced nanoparticles modeled as dipoles \cite{kathmann2018limitations,tervo2020comparison}. These studies suggest that the kinetic theory provides reasonable estimates of the total radiative conductivity but is less accurate in capturing its spectral distribution. In contrast, Kathmann et al. \cite{kathmann2018limitations} concluded that the kinetic theory is a severely limited framework for predicting both the spectral and total radiative conductivity in systems of spaced dipoles.

The absence of a comprehensive FE-based model has hindered the interpretation of thermal conductivity measurements in polaritonic nanofilms \cite{tranchant2019two,wu2020enhanced,wu2022observation} and nanowires \cite{pan2023remarkable}. This work addresses this knowledge gap by leveraging the FE-based, numerically exact discrete system Green's function (DSGF) method \cite{walter:2022near,correa2024dsgf}  to model thermal electromagnetic transport in solid-state systems. Specifically, the DSGF method is applied to predict the radiative thermal conductivity of SiO\textsubscript{2} nanowires with diameters of 66 nm and 132 nm in the diffusive regime, and the results are compared with those obtained using the kinetic theory. It is shown that, while the kinetic theory predicts a maximum radiative conductivity on the order of the bulk phonon conductivity of SiO\textsubscript{2}, the DSGF method yields a maximum value on the order of $10^{-2} $ W m$^{-1}$ K$^{-1}$. This discrepancy can be mitigated by introducing an effective mean free path in the kinetic theory, which improves predictions of the total radiative conductivity but does not resolve the significant differences in the spectral distribution. This work highlights the importance of FE-based frameworks for predicting thermal electromagnetic transport in solid-state systems, enabling the interpretation of experimental data and providing a foundation for the design of phonon polariton-based technologies for thermal management in micro/nanoelectronic devices.

\section*{Results}

Within the framework of FE, the DSGF method discretizes thermal sources into cubic subvolumes and evaluates the system Green’s function for all subvolume pairs \cite{walter:2022near}. The subvolume side lengths must be smaller than both the vacuum and material wavelengths to justify the assumption of spatially uniform electric fields and Green's functions within each subvolume. The DSGF method has been validated through comparison with experimental measurements of near-field radiative heat transfer between SiC subwavelength membranes \cite{tang2024corner}. 

\begin{figure}
	\centering
	\includegraphics[width=1\linewidth]{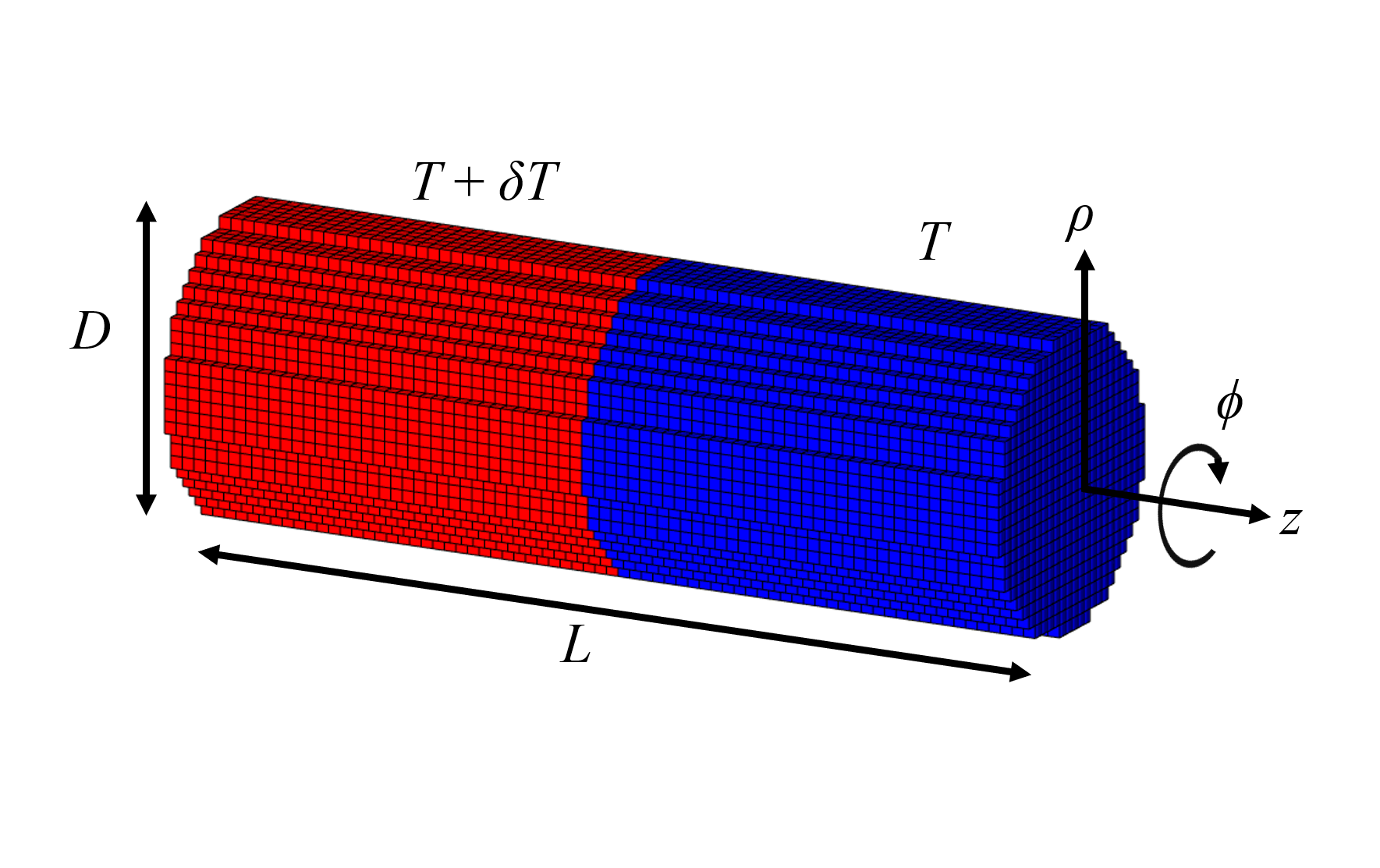}
	\caption{\label{fig_DSGF}Thermal electromagnetic transport in the diffusive regime along a SiO$_2$ nanowire of diameter $D$ and lenght $L$. In the schematic, the nanowire diameter is 66 nm and the diffuse regime is reached for a length of 200 nm. The nanowire is discretized into 32,184 uniform cubic subvolumes. One half of the nanowire is maintained at temperature $T +\delta T$, while the other half is at temperature $T$.}
\end{figure}

The system under study consists of a SiO$_2$ nanowire of length $L$  and diameter $D$ discretized into $N$ subvolumes and surrounded by free space (Fig. \ref{fig_DSGF}). The frequency-dependent dielectric function of SiO$_2$ was obtained from Ref. \cite{czapla2019thermal} and is shown in \textit{SI Appendix}, Fig. S1.  The nanowire diameter is sufficiently small to consider one-dimensional thermal transport along the axial direction. The nanowire is artificially divided into two isothermal sections of equal volumes $V_A$ and $V_B$. The hot section is at a temperature $T + \delta T$, whereas the cold section is at temperature $T$. In the limit that $ \delta T \rightarrow 0$, the total radiative thermal conductivity along the \textit{z}-direction is calculated as follows \cite{tervo2019thermal,walter2025generalized}:

\begin{equation}
	\label{eq:conductivity_DSGF}
	\begin{split}
		\kappa_{\text{DSGF}}(T)
		= \int_{0}^{\infty}\frac{1}{2\pi A_c}\left[\sum_{i\in V_A}\sum_{j\in V_B}\mathcal{T}_{i j}(\omega)
		(\mathbf{r}_j-\mathbf{r}_i)\cdot \hat{\mathbf{z}}\right] \\
		\times \left[\frac{\partial {\Theta}(\omega,T')}{\partial T}\right]_{T'=T}d\omega
	\end{split}
\end{equation}

\noindent where $A_c$ is the nanowire cross-sectional area, $\mathbf{r}_i$ and $\mathbf{r}_j$ are the center points of subvolumes $i$ and $j$, and $\Theta = \hbar \omega/(e^{\hbar \omega/{k_B T}} - 1)$ is the mean energy of an electromagnetic state with $\hbar$ as the reduced Planck constant and $k_B$ the Boltzmann constant. The spectral transmission coefficient between subvolumes \textit{i} and \textit{j}, $\mathcal{T}_{ij}$, is calculated from the system Green's functions as described in the \textit{Materials and Methods} section.

The DSGF-predicted total radiative conductivity in the diffusive regime for $D$ = 66 nm and 132 nm is shown as a function of temperature in Figs. 2\textit{A} and 2\textit{B}, respectively. For both diameters, the results were obtained using a nanowire length $L$ = 200 nm, as increasing $L$ does not produce a perceptible change in the radiative conductivity. For example, increasing the nanowire length to 1 $\mu$m results in a relative difference in radiative conductivity of less than 2\% as shown in \textit{SI Appendix}, Fig. S2. The 66 nm and 132 nm nanowires were discretized into 32,184 and 32,568 uniform subvolumes, respectively. A convergence analysis of the DSGF method is provided in \textit{SI Appendix}, section S2. 

The radiative conductivity is modest and takes values of 0.0264 and 0.0165 W m$^{-1}$ K$^{-1}$ at 400 K for \textit{D} values of 66 nm and 132 nm, respectively. The observed increase in conductivity with decreasing nanowire characteristic dimension is in qualitative agreement with previous studies of radiation conduction in SiO$_2$ nanofilms based on the kinetic theory \cite{chen2005surface}. 

\begin{figure*}[h!] 
	\includegraphics[width=1.\linewidth]{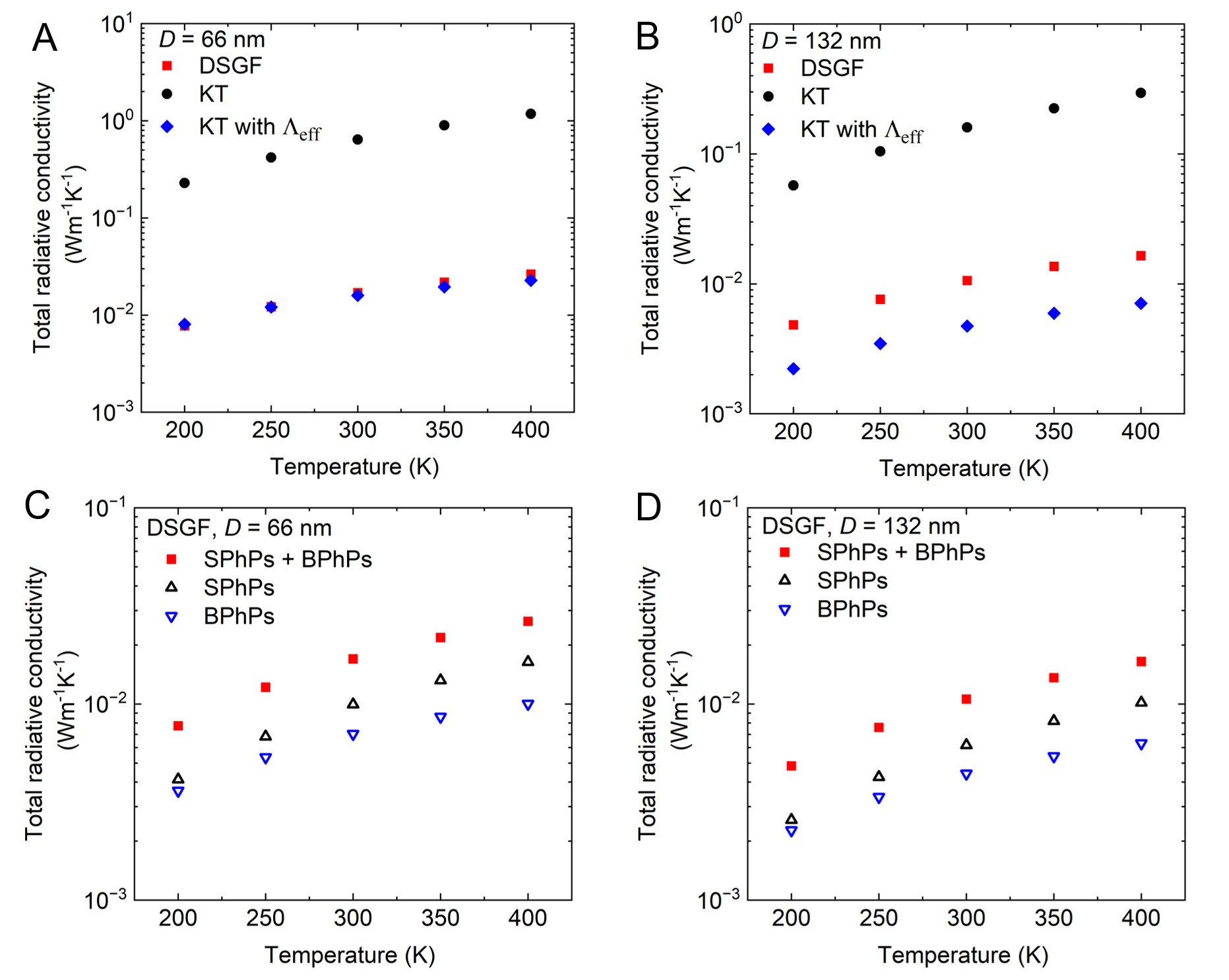}
	\caption{Total radiative conductivity along SiO$_2$ nanowires in the diffusive regime as a function of temperature calculated using the DSGF method, the conventional kinetic theory (KT) and the kinetic theory with effective mean free path (KT with $\Lambda_{\text{eff}}$)  for: (A) $D$ = 66 nm  and (B) $D$ = 132 nm. Contributions of SPhPs and BPhPs to the total radiative conductivity as a function of temperature predicted with the DSGF method for: (C) $D$ = 66 nm  and (D) $D$ = 132 nm.}
	\label{fig:total_conductivity}
\end{figure*}

SPhPs exist within the two Reststrahlen bands of SiO$_2$, spanning angular frequency ranges of 8.70-9.67$\times 10^{13}$ rad/s and 2.04-2.33$\times 10^{14}$ rad/s. Within these bands, SiO$_2$ exhibits a metallic behavior, with a negative real part of its dielectric function, thereby preventing wave propagation within the nanowire volume. Outside the Reststrahlen bands, where the real part of the dielectric function of SiO$_2$ is positive, radiative transport is mediated by bulk phonon polaritons (BPhPs), which are electromagnetic waves propagating within the volume of the nanowire \cite{li2024phonon}. Figs. 2\textit{C} and 2\textit{D} present DSGF predictions of the individual contributions of SPhPs and BPhPs to the total radiative conductivity as a function of temperature, showing that SPhPs are the largest contributor to the radiative conductivity for both nanowire diameters. 

DSGF results are compared with those obtained using the conventional approach for predicting radiation conduction, which combines the kinetic theory with dispersion relations derived from Maxwell's equations. The total radiative conductivity in the diffusive regime for a nanowire, as predicted by the kinetic theory, is given by \cite{tervo2019photonic,guo2021heat}:

\begin{gather} 
	\kappa_{\text{KT}} (T) = \int_{0}^{\infty}\frac{ \Lambda }{\pi A_c} \left[\frac{\partial {\Theta}(\omega,T')}{\partial T}\right]_{T'=T}d \omega,
\end{gather}

\noindent where $\Lambda$ is the radiation mean free path obtained from the dispersion relation. Starting from Maxwell's equations in cylindrical coordinates and assuming axial wave propagation along an infinitely long cylinder, the following dispersion relation is derived for the zeroth-order mode in transverse magnetic polarization \cite{sarid2010modern}:

\begin{gather}
	\frac{\varepsilon_1}{\gamma_1}\frac{I_1(\gamma_1D/2)}{I_0(\gamma_1D/2)} = 	\frac{\varepsilon_2}{\gamma_2}\frac{K_1(\gamma_2D/2)}{K_0(\gamma_2D/2)}, 
\end{gather}

\noindent where $I_n$ and $K_n$ are respectively the modified Bessel functions of the first and second kind of order $n$, $\gamma_m=\sqrt{k_z^2-\varepsilon_mk_0^2}$ is the radial wave vector in medium $m$ with dielectric function $\varepsilon_m$, $k_z$ is the axial wave vector, and $k_0$ is the vacuum wave vector. 

The dispersion relation is solved numerically for $k_z$ following the methodology described in Ref. \cite{pan2023remarkable}. The radiation mean free path is then obtained from the axial wave vector $k_z$ as $\Lambda=1/[2\text{Im}(k_z)]$. The real and imaginary parts of $k_z$ for nanowire diameters of 66 nm and 132 nm are shown in Figs. 3\textit{A} and 3\textit{B}. Outside the Reststrahlen bands, the real and imaginary parts of $k_z$ are the same as the material light calculated as $k=\sqrt{\varepsilon_{\text{SiO$_2$}}}k_0$. This behavior further confirms that BPhPs existing outside the Reststrahlen bands are electromagnetic modes propagating within the nanowire volume.

\begin{figure*}[t!]
	\includegraphics[width=1.\linewidth]{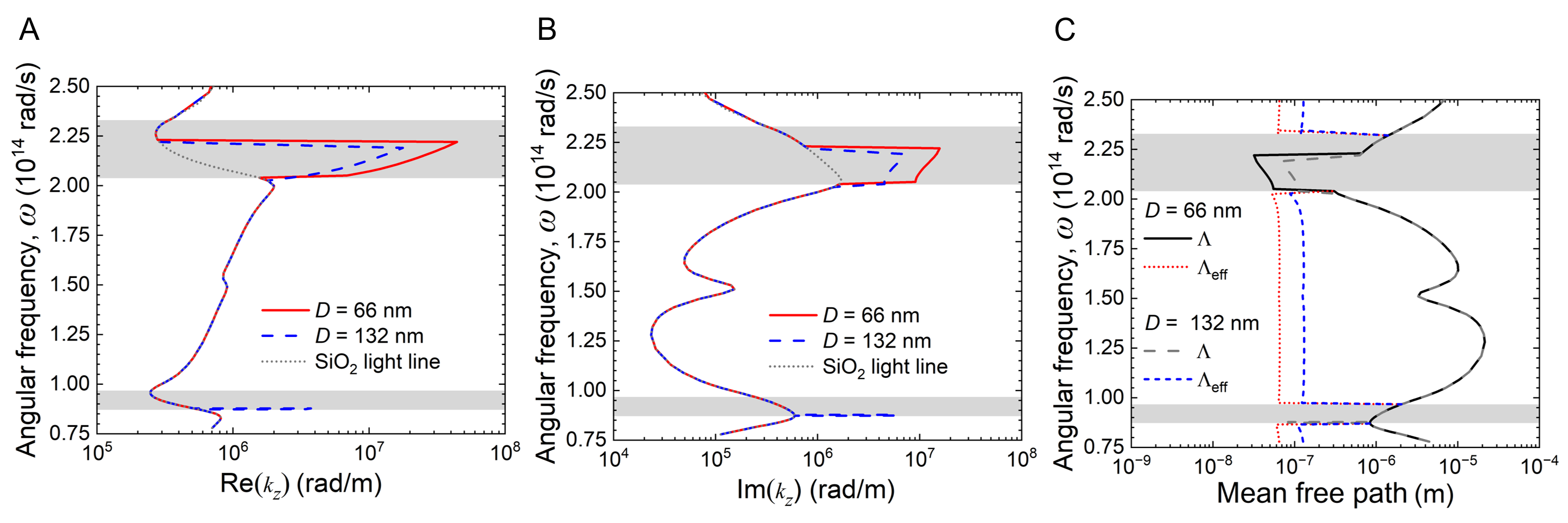}
	\caption{Dispersion relations for SiO$_2$ nanowires of infinite length and diameters of 66 nm and 132 nm:  (A) Angular frequency $\omega$ as a function of the real part of the axial wave vector $\text{Re}(k_z)$ and (B) angular frequency $\omega$ as a function of the imaginary part of the axial wave vector $\text{Im}(k_z)$. In both panels (A) and (B), the light line in SiO$_2$ is plotted. (C) Spectral mean free path for $D$ = 66 nm and 132 nm. The mean free path $\Lambda$ is calculated as $[2\text{Im}(k_z)]^{-1}$, whereas the effective mean free path $\Lambda_{\text{eff}}$ is calculated using Eq. \ref{eq:effec_MFP} outside the Reststrahlen bands only. In all panels, the Reststrahlen bands are identified by gray-shaded boxes.}
	\label{fig:dispersion_relations}
\end{figure*}

The radiative conductivity predicted by the kinetic theory is compared with DSGF results in Figs. 2\textit{A} and 2\textit{B}. It can be seen that the kinetic theory significantly overestimates the radiative conductivity for both nanowire diameters. For example, for a nanowire diameter of 66 nm at 400 K, the radiative conductivity is predicted to be 1.18 W m$^{-1}$ K$^{-1}$, which is approximately 45 times larger than the value obtained from FE-based DSGF simulations and similar to the bulk phonon thermal conductivity of SiO$_2$ of approximately 1.4 W m$^{-1}$ K$^{-1}$. 

As shown in Fig. 3\textit{C}, the radiation mean free path is the largest outside Reststrahlen bands, reaching values slightly exceeding 10 $\mu$m, whereas within the Reststrahlen bands it attains maximum values on the order of 1 $\mu$m. This indicates that the radiative conductivity predicted by the kinetic theory is dominated by BPhPs propagating within the nanowire volume, rather than by SPhPs traveling along the nanowire surface as observed with DSGF predictions. This distinction is further illustrated in Figs. 4\textit{A} and 4\textit{B}, which compare the spectral radiative conductivity at 400 K from the DSGF and kinetic theory for nanowire diameters of 66 nm and 132 nm, respectively. 

It is well established that the phonon thermal conductivity is reduced in nanostructures due to increased boundary scattering \cite{zhang2007nano}. A similar size effect applies to photons propagating within the volume of nanostructures; here, these photons correspond to BPhPs. Consequently, an effective mean free path, $\Lambda_{\text{eff}}$, should be used in the kinetic theory outside the Reststrahlen spectral bands to account for the size effect. Using Matthiessen's rule, the effective radiation mean free path is defined as:

\begin{gather}\label{eq:effec_MFP}
	\frac{1}{\Lambda_{\text{eff}}} = \frac{1}{\Lambda} + \frac{1}{\Lambda_w}
\end{gather}

\noindent where $\Lambda$ is the mean free path calculated using the nanowire dispersion relation, and $\Lambda_w$ is the reduced mean free path within the nanowire. This latter quantity is calculated as follows \cite{zhang2007nano}: 

\begin{gather}\label{eq:size_effect_MFP}
	\Lambda_w = \Lambda\left(\frac{1}{\text{Kn}} - \frac{1}{4\text{Kn}^2}\right), 
\end{gather}
\noindent where $\text{Kn}= \Lambda/D$ is the Knudsen number. Eq. \ref{eq:size_effect_MFP} is valid for $\text{Kn} > 5$, a condition satisfied for all frequencies outside the Reststrahlen bands. The effective radiation mean free path is shown in Fig. 3\textit{C}, where it is reduced by approximately one to two orders of magnitude. 

The effective radiation mean free path $\Lambda_{\text{eff}}$ outside the Reststrahlen bands is used to compute a modified radiative conductivity from the kinetic theory. Within the Reststrahlen bands, the radiation mean free path $\Lambda$ is used, as SPhPs do not experience size effects, since they travel along the interface between free space and SiO$_2$. 

The radiative conductivity with effective mean free path, shown in Figs. 2\textit{A} and 2\textit{B}, approaches that from the DSGF predictions. The agreement is remarkable for a nanowire diameter of 66 nm; at 400 K, the total radiative conductivity predicted by the kinetic theory with effective mean free path is 0.0227 W m$^{-1}$ K$^{-1}$, which is only 14\% lower than the DSGF-predicted conductivity. For the 132 nm diameter, the agreement is not as strong. At 400 K, the radiative conductivity predicted by the kinetic theory with effective mean free path is 0.00707 W m$^{-1}$ K$^{-1}$, which is slightly less than half of the value predicted with the DSGF.  

When the size effect is accounted for in the kinetic theory, the radiative conductivity becomes dominated by SPhPs rather than BPhPs. This is evident from Fig. 3\textit{C}, where the effective mean free path outside the Reststrahlen bands is smaller than that within the Reststrahlen bands. This behavior is also clearly reflected in the spectral distributions of radiative conductivity at 400 K shown in Figs. 4\textit{A} and 4\textit{B}. It is also noteworthy that,  even for the 66 nm diameter nanowire where the total radiative conductivity predicted by the DSGF is in excellent agreement with that obtained from the kinetic theory with effective mean free path, the spectral distributions of radiative conductivity differ significantly both within and outside the Reststrahlen bands.

\begin{figure}
	\centering
	\includegraphics[width=0.8\linewidth]{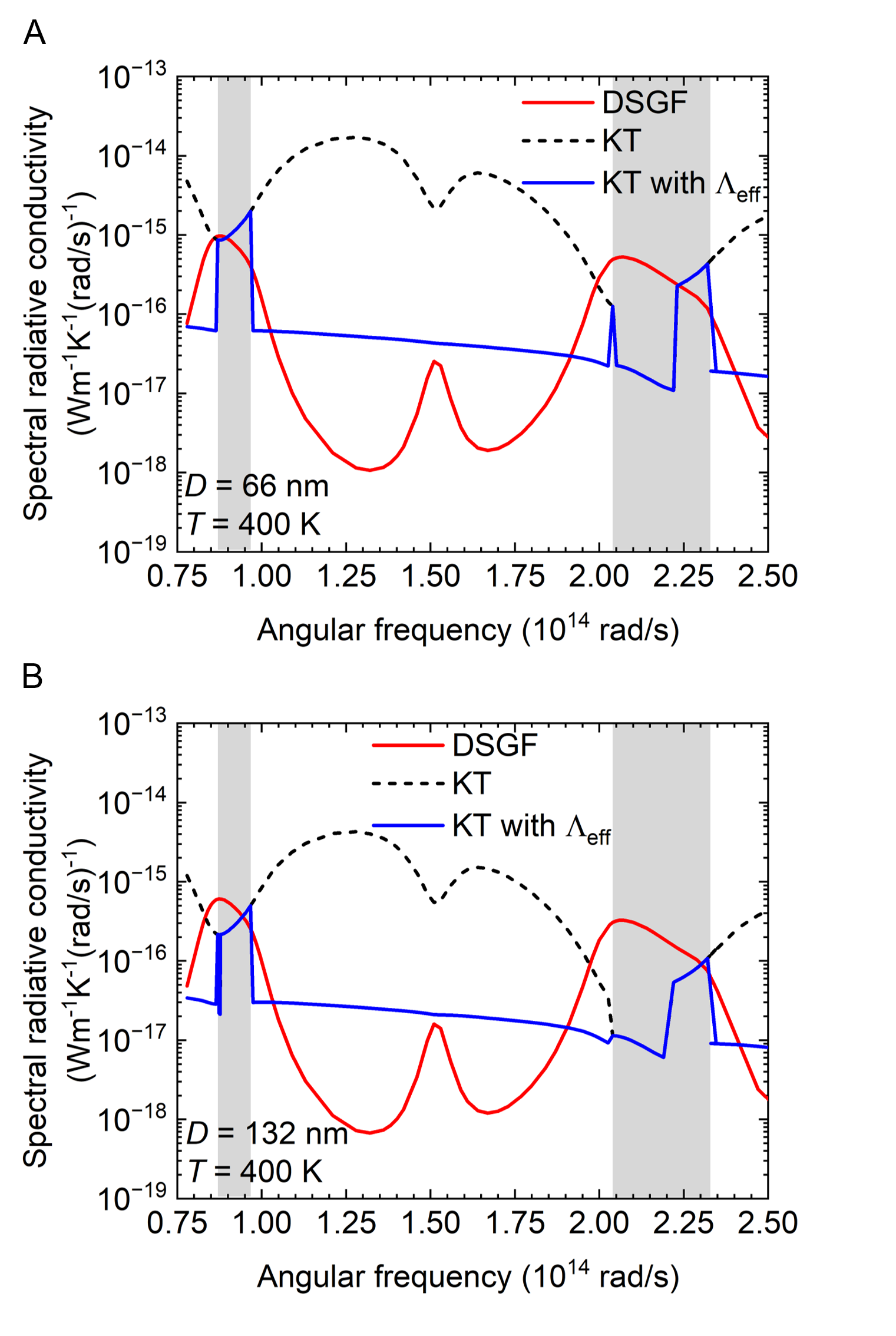}
	\caption{\label{KT_vs_DSGF} Spectral radiative conductivity along SiO$_2$ nanowires in the diffusive regime at 400 K calculated using the DSGF method, the conventional kinetic theory (KT) and the kinetic theory with effective mean free path (KT with $\Lambda_{\text{eff}}$)  for: (A) $D$ = 66 nm  and (B) $D$ = 132 nm. In both panels, the Reststrahlen bands are identified by gray-shaded boxes.}
\end{figure}

\section*{Discussion and Conclusion}

FE-based simulations of thermal electromagnetic transport in solid-state systems should also be validated against experimental data. To date, only Pan \textit{et al.} \cite{pan2023remarkable} have measured the thermal conductivity along 3C-SiC nanowires. The authors reported that radiation has a negligible impact on the thermal conductivity of bare SiC nanowires with a diameter of 66 nm. This observation is in qualitative agreement with the DSGF results presented here, which indicate that SPhPs and BPhPs have a negligible contribution to thermal transport, on the order of $10^{-2} $ W m$^{-1}$ K$^{-1}$, which is two orders of magnitude smaller than the bulk phonon thermal conductivity of SiO$_2$. 

In addition, using the kinetic theory with an effective mean free path, a negligibly small radiative conductivity of 0.051 W m$^{-1}$ K$^{-1}$ at 400 K is predicted for a 66 nm diameter, 47.5 $\mu$m long SiC nanowire, as shown in \textit{SI Appendix}, Fig. S4. In contrast, a radiative conductivity of 1.01 W m$^{-1}$ K$^{-1}$ is predicted using the conventional kinetic theory. These results suggest that the kinetic theory with an effective mean free path, as implemented here and benchmarked against the DSGF method, provides reasonable estimates of the total radiative conductivity consistent with experimental observations.

A significant radiative contribution to the thermal conductivity, for example, 5.8 W m$^{-1}$ K$^{-1}$ at 300 K, was reported in Ref. \cite{pan2023remarkable} when the ends of the SiC nanowires are coated with Au. However, such systems consisting of SiC nanowires with lengths ranging from approximately 10 $\mu$m to 50 $\mu$m and micrometer-scale Au coatings cannot currently be modeled using the DSGF method due to computational limitations. The FE-based DSGF method is a comprehensive computational framework that, in principle, enables full-wave simulations of thermal electromagnetic transport in complex multiscale solid-state systems. In practice, the maximum number of subvolumes that can be simulated is on the order of 40,000 for 1 TB of RAM, as detailed in Ref. \cite{correa2024dsgf}. The absolute value of the real part of the dielectric function of SiC and Au is significantly larger than that of  SiO$_2$, requiring smaller subvolumes due to shorter material wavelengths in these materials. In this work, 32,184 subvolumes were required to obtain converged results for a 200 nm long, 66 nm diameter SiO$_2$ nanowire. Modeling a bare SiC nanowire of the same size would require at least twice the aforementioned number of subvolumes, which is currently impractical.

It is worth noting that larger systems involving near-field radiative heat transfer between subwavelength membranes of SiC, SiO$_2$, and SiN have been successfully modeled using the DSGF method \cite{tang2024corner,mccormack2026near}. For near-field radiative heat transfer, where most of the energy transfer occurs near the gap spacing, computational costs can be significantly reduced using nonuniform discretization, with finer subvolumes near the gap and coarser ones farther away. In contrast, for radiation conduction, all regions of the system contribute significantly to the radiative conductivity, making nonuniform discretization ineffective. 

In summary, this work demonstrated that the kinetic theory previously used to predict radiation conduction along polaritonic nanowires significantly overestimates the radiative thermal conductivity. For a 66 nm diameter SiO$_2$ nanowire in the diffusive regime, this overestimation is approximately two orders of magnitude at 400 K. Improved agreement in the total radiative conductivity between full-wave FE-based simulations and kinetic theory is achieved when accounting for the reduced mean free path of BPhPs propagating within the nanowire volume. However, the spectral radiative conductivity remains poorly predicted by the kinetic theory, even when an effective mean free path is used. Enhancements to the publicly available DSGF solver \cite{correa2024dsgf} could enable full-wave simulations of larger-scale systems, facilitating the development of technologies that leverage phonon polaritons for efficient heat dissipation in micro/nanoelectronic devices.



\section*{Materials and methods}

	\subsection*{Spectral transmission coefficient for the discrete system Green's function (DSGF) method} 
	
	The spectral transmission coefficient in \eqref{eq:conductivity_DSGF} between subvolumes $i$ and $j$ occupying respective volumes $\Delta V_i$ and $\Delta V_j$  is calculated as follows \cite{walter:2022near}: 
	\begin{multline}
		\label{eq:trans_coeff}
		\mathcal{T}_{ij}(\omega)=4k_0^4\Delta V_i \Delta V_j\text{Im}[\varepsilon(\mathbf{r}_i,\omega)]\text{Im}[\varepsilon(\mathbf{r}_j,\omega)]\times \\ \text{Tr}[
		\mathbf{\bar{\bar{G}}}
		(\mathbf{r}_i,\mathbf{r}_j,\omega)
		\mathbf{\bar{\bar{G}}}^\dagger(\mathbf{r}_i,\mathbf{r}_j,\omega)],  
	\end{multline}
	where $\varepsilon$ is the dielectric function, $^\dagger$ is the conjugate transpose operator, $\text{Tr}$ is the trace operator, and $\bar{\bar{\mathbf{G}}}(\mathbf{r}_i,\mathbf{r}_j,\omega)$ is the monochromatic system Green's function relating subvolumes $i$ and $j$. The system Green's functions between $N$ subvolumes in the nanowire are calculated by solving the following system of linear equations \cite{walter:2022near}: 
	
	\begin{multline}
		\label{eq:SGF_matrix}
		\left\{
		\begin{bmatrix}
			\mathbf{\bar{\bar{I}}} & 0  & 0 \\
			0 & \ddots  & 0 \\
			0 & 0  & \mathbf{\bar{\bar{I}}}
		\end{bmatrix} 
		-k_0^2
		\begin{bmatrix}
			\mathbf{\bar{\bar{G}}}_{11}^0 & \ldots  & \mathbf{\bar{\bar{G}}}_{1N}^0 \\
			\vdots & \ddots  & \vdots \\
			\mathbf{\bar{\bar{G}}}_{N1}^0 & \ldots  & \mathbf{\bar{\bar{G}}}_{NN}^0
		\end{bmatrix} 
		\begin{bmatrix}
			\alpha_{1}^{(0)} & 0  & 0 \\
			0 & \ddots  & 0 \\
			0 & 0  & \alpha_{N}^{(0)}
		\end{bmatrix}
		\right\} \\
		\times
		\begin{bmatrix}
			\mathbf{\bar{\bar{G}}}_{11} & \ldots  & \mathbf{\bar{\bar{G}}}_{1N} \\
			\vdots & \ddots  & \vdots \\
			\mathbf{\bar{\bar{G}}}_{N1} & \ldots  & \mathbf{\bar{\bar{G}}}_{NN}
		\end{bmatrix} 
		= \begin{bmatrix}
			\mathbf{\bar{\bar{G}}}_{11}^0 & \ldots  & \mathbf{\bar{\bar{G}}}_{1N}^0 \\
			\vdots & \ddots  & \vdots \\
			\mathbf{\bar{\bar{G}}}_{N1}^0 & \ldots  & \mathbf{\bar{\bar{G}}}_{NN}^0
		\end{bmatrix} , 
	\end{multline}
	where the subscripts $ij$ refer to lattice locations $\mathbf{r}_i$ and $\mathbf{r}_j$, $\alpha^{(0)}_i={\Delta}V_i [\varepsilon(\mathbf{r}_i,\omega) -1]$ is the bare polarizability of subvolume $i$, $\mathbf{\bar{\bar{I}}}$ is the unit dyadic, and $\mathbf{\bar{\bar{G}}}_{ij}^0$ is the free space Green's function between subvolumes $i$ and $j$. 
	
	The discretized free-space Green's function in vacuum is given by: 
	\begin{multline}
		\label{eq:free_space_GF_discretized}  
		\bar{\bar{\mathbf{G}}}^0(\mathbf{r}_i,\mathbf{r}_j,\omega) = \frac{e^{(ik_0r_{ij})}}{4\pi r_{ij}} \times \\ \left[ \left( 1 - \frac{1}{(k_0 r_{ij})^2}+ \frac{i}{(k_0 r_{ij})}\right) \bar{\bar{\mathbf{I}}} - \left( 1 - \frac{3}{(k_0 r_{ij})^2} + \frac{3i}{(k_0 r_{ij})}\right) \mathbf{\hat{r}}_{ij}\mathbf{\hat{r}}_{ij}^\dagger \right], \\ j\neq i,
	\end{multline}
	where $r_{ij} = |\mathbf{r}_i-\mathbf{r}_j|$ and  $\mathbf{\hat{r}}_{ij} = \frac{\mathbf{r}_i-\mathbf{r}_j}{|\mathbf{r}_i-\mathbf{r}_j|}$. 
	
	At the singularity where $\mathbf{r}_i=\mathbf{r}_j$, the discretized free-space Green's function is derived from the principal value method \cite{yaghjian1980electric} and is calculated as: 
	\begin{multline}
		\label{eq:free_space_GF_discretized}  
		\bar{\bar{\mathbf{G}}}^0(\mathbf{r}_i,\mathbf{r}_j,\omega) = \frac{\bar{\bar{\mathbf{I}}}}{3 \Delta V_j k_0^2}\{2[e^{(ik_0 a_{j})}(1-i a_j k_0)-1]-1\}, \quad j= i,
	\end{multline}
	where $a_{j} = (3 \Delta V_j/(4\pi))$.

\bibliographystyle{elsarticle-num}
\bibliography{bib_cond.bib}



\section*{Supplementary material}

# Supporting Information

## Radiation conduction in polaritonic nanowires

Lívia Corrêa McCormack[1] and Mathieu Francoeur[1*]

Department of Mechanical Engineering, The University of Utah, Salt Lake City, UT 84112, USA

[*]Corresponding author: mfrancoeur@mech.utah.edu

## S1. DIELECTRIC FUNCTION OF $SiO_2$

The real and imaginary parts of the dielectric function of $SiO_2$, obtained from Ref. [1], is shown in Fig. S1.

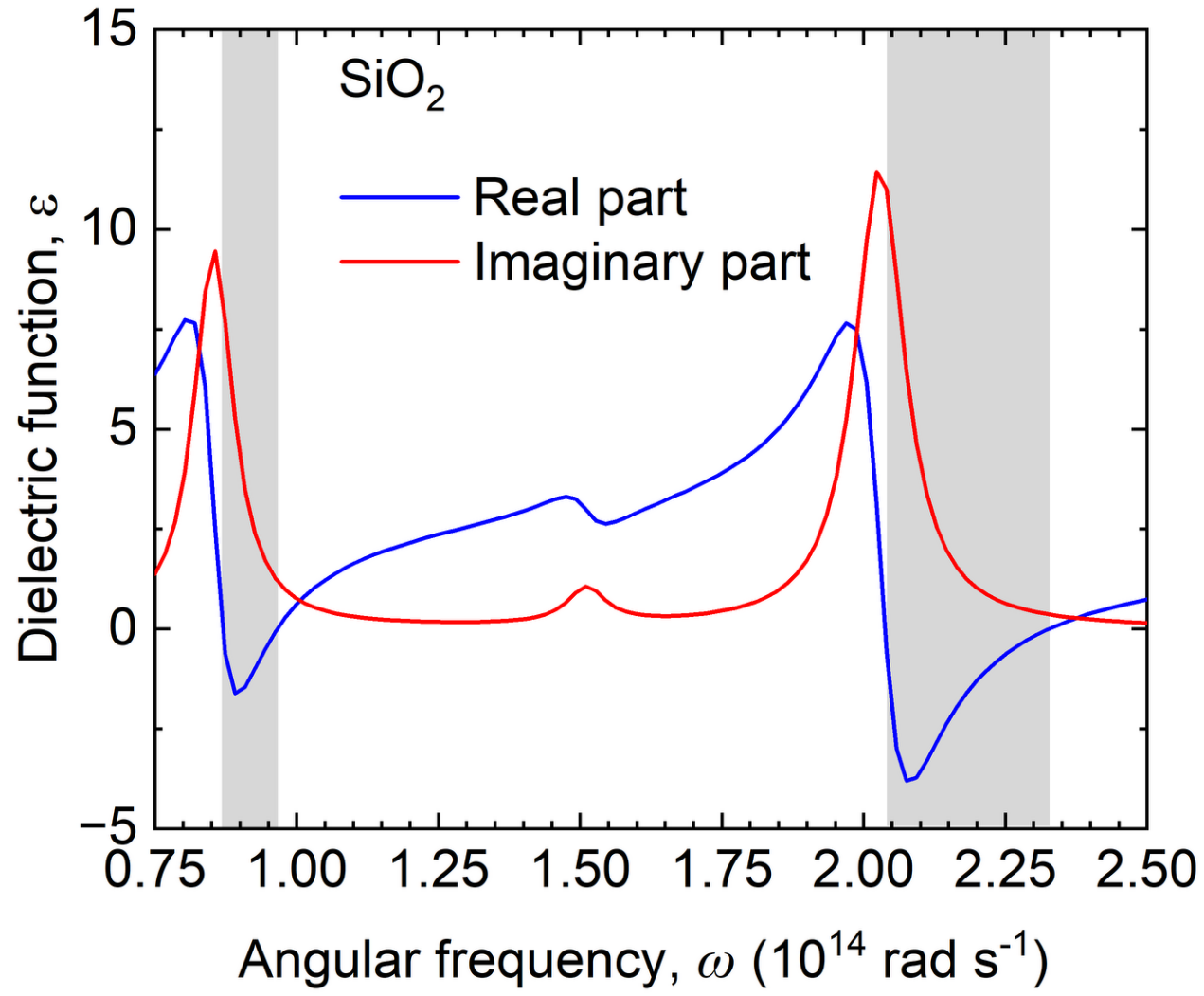


**Fig. S1.** Real and imaginary parts of the dielectric function of $SiO_2$. The Reststrahlen bands are identified by the gray-shaded boxes.

## S2. CONVERGENCE OF THE DISCRETE SYSTEM GREEN'S FUNCTION (DSGF) METHOD

Fig. S2 shows the relative difference in total radiative conductivity calculated using the discrete system Green's function (DSGF) method for a 66 nm diameter $SiO_2$ nanowire as a function of nanowire length. The relative difference in total radiative conductivity is less than 1.5% when the nanowire length is increased from 150 nm to 200 nm. Further increasing the length to 250 nm results in a relative difference of less than 1%. To balance accuracy and computational cost, a length of 200 nm was selected, as further increases did not produce perceptible changes in the radiative conductivity.

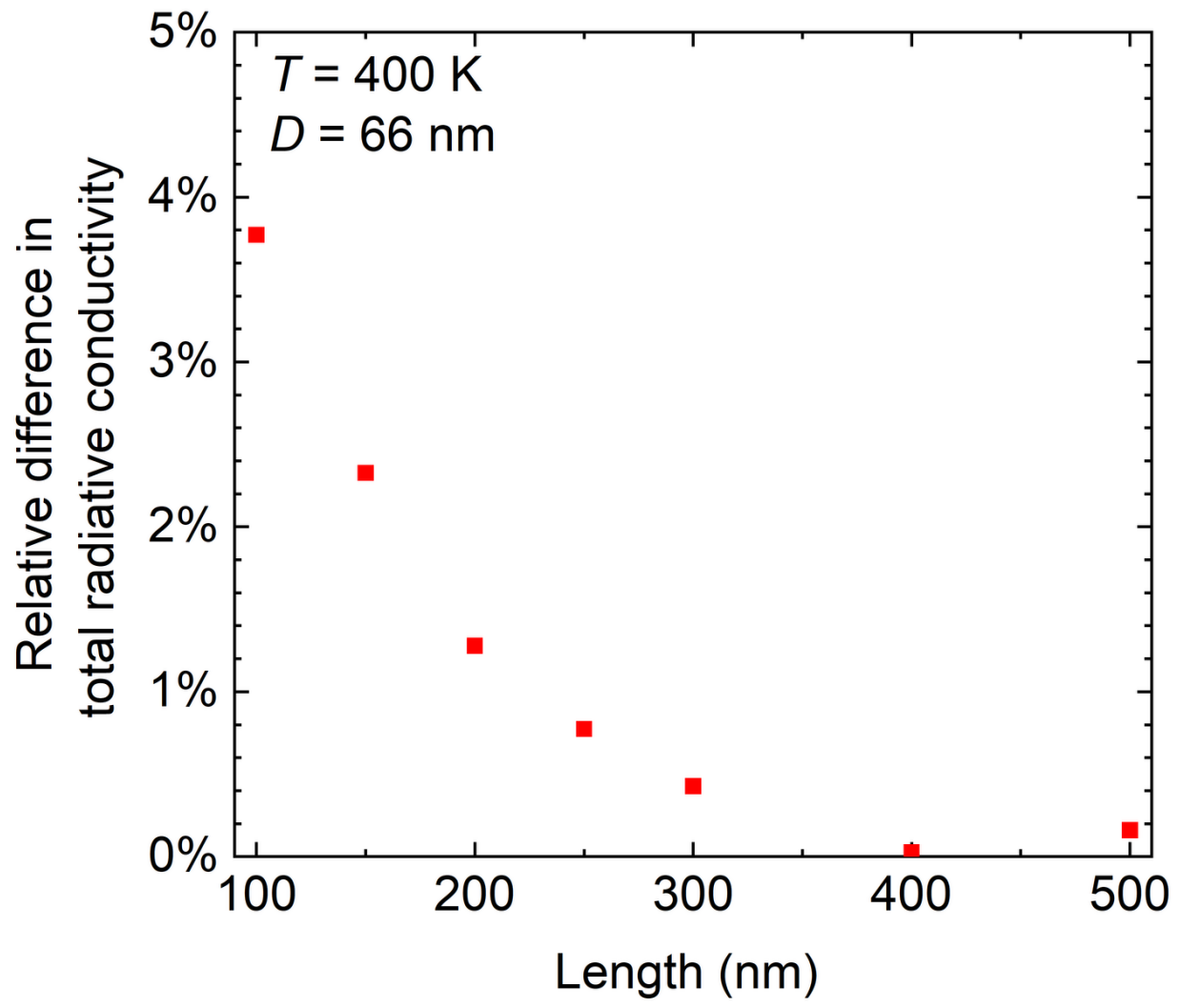


**Fig S2.** Relative difference in total radiative conductivity as a function of the $SiO_2$ nanowire length. The diameter and temperature are fixed at 66 nm and 400 K, respectively.

Fig. S3 shows the relative difference in radiative conductivity calculated with the DSGF method at a frequency of $8.73\times10^{13}$ rad/s, located within the low-frequency Reststrahlen band of $SiO_2$, as a function of the subvolume side length $L_{sub}$. The nanowire diameter and length are fixed at 66 nm and 200 nm, respectively. The relative difference in radiative conductivity for subvolume side lengths of 2.77 nm and 2.98 nm, corresponding to total numbers of subvolumes $N$ of 32,184 and 26,112, respectively, is 7.21%. A subvolume side length of 2.77 nm was selected to balance accuracy and computational cost, which is approximately 658 core-hours per frequency.

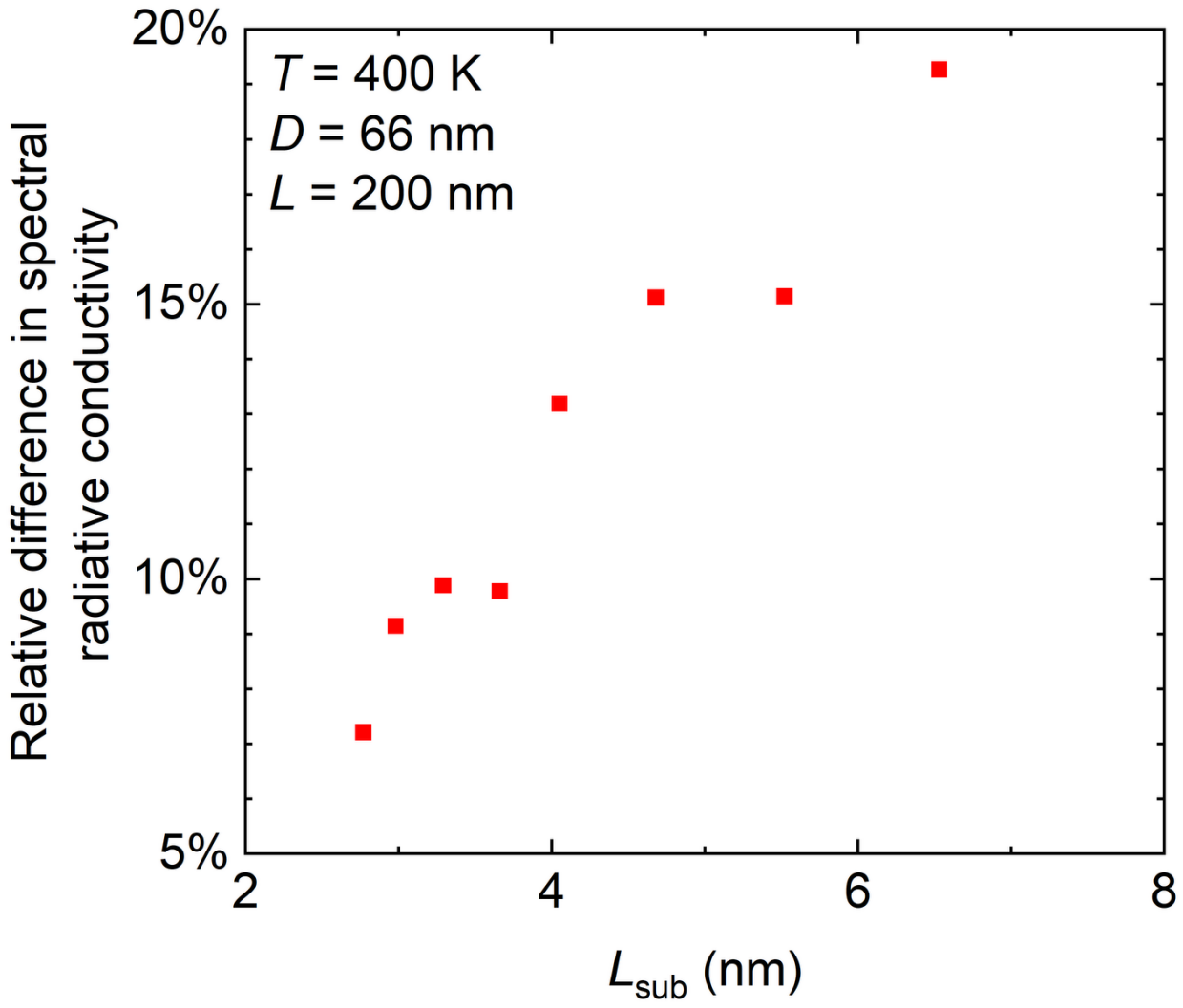


**Fig. S3.** Relative difference in radiative conductivity a $SiO_2$ nanowire at a frequency of $8.73\times10^{13}$ rad/s as a function of the subvolume side length $L_{sub}$. The diameter, length and temperature are fixed at 66 nm, 200 nm, and 400 K, respectively.

A similar convergence analysis was also performed for the 132 nm diameter $SiO_2$ nanowire.

## S3. RADIATIVE THERMAL CONDUCTIVITY OF A SiC NANOWIRE

The radiative thermal conductivity of a SiC nanowire with diameter $D$ = 66 nm and length $L$ = 47.5 μm, corresponding to the dimensions investigated experimentally in Ref. [2], is predicted as a function of temperature using both the conventional kinetic theory [3] and the kinetic theory with an effective mean free path. Since the diffusive regime is not reached in a 47.5 μm-long SiC nanowire, the following expression for the radiative conductivity of a finite-length nanowire is used [3]:

$$\kappa_{\mathrm{KT}}(T) = \int_0^{\infty} \frac{\Lambda}{\pi A_c}\left(1 - \frac{\psi(0)}{\lambda}\right)\left[\frac{\partial \Theta(\omega, T')}{\partial T}\right]_{T'=T} d\omega, \tag{S1}$$

where $\Lambda$ is the radiation mean free path obtained from the SiC nanowire dispersion relation, $A_c$ is the nanowire cross-sectional area, $\Theta$ is the mean energy of an electromagnetic state, $\lambda = L/\Lambda$ is the ratio of the nanowire length to the mean free path, and $\psi(0) = 1 - e^{-\lambda}$.

The radiative conductivity predicted from Eq. (S1), with and without the effective mean free path outside the Reststrahlen band of SiC, is shown in Fig. S4.

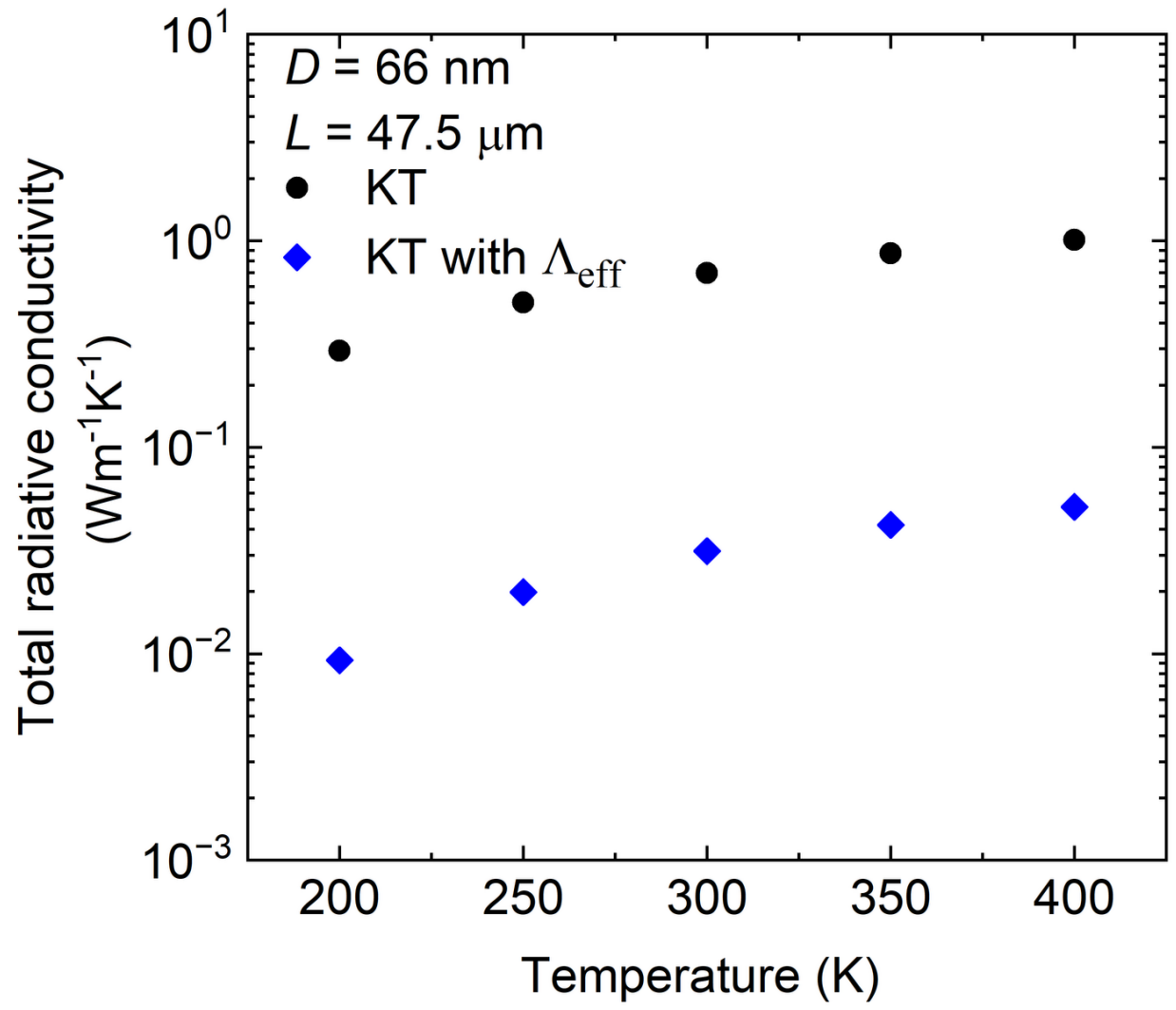


**Fig. S4.** Total radiative conductivity along a SiC nanowire of diameter $D$ = 66 nm and length $L$ = 47.5 μm as a function of temperature calculated using the conventional kinetic theory (KT) and the kinetic theory with effective mean free path (KT with $\Lambda_{\mathrm{eff}}$).










\end{document}